\documentstyle[epsfig]{article}
\newcommand{\be}{\begin{equation}} 
\newcommand{\ee}{\end{equation}} 
\newcommand{\bea}{\begin{eqnarray}} 
\newcommand{\eea}{\end{eqnarray}} 
\newcommand{\equ}[1]{eq.~(\ref{#1})}

\catcode`\_=11 
\def\c_ncel#1#2{\ooalign{\lower.2ex\hbox{$\hfil#1\mkern.7mu/\hfil$}\crcr$#1#2$}} 
\def\dslash{\mathord{\mathpalette\c_ncel d}} 
\catcode`\_=8 
\catcode`\_=11 
\def\c_ncel#1#2{\ooalign{\lower.2ex\hbox{$\hfil#1\mkern.7mu/\hfil$}\crcr$#1#2$}} 
\def\aslash{\mathord{\mathpalette\c_ncel a}} 
\catcode`\_=8 
\catcode`\_=11 
\def\c_ncel#1#2{\ooalign{\lower.2ex\hbox{$\hfil#1\mkern.7mu/\hfil$}\crcr$#1#2$}} 
\def\bslash{\mathord{\mathpalette\c_ncel b}} 
\catcode`\_=8 
\catcode`\_=11 
\def\c_ncel#1#2{\ooalign{\lower.2ex\hbox{$\hfil#1\mkern.7mu/\hfil$}\crcr$#1#2$}} 
\def\cslash{\mathord{\mathpalette\c_ncel c}} 
\catcode`\_=8 

\def\ubar{{\overline u}}
\def\vbar{{\overline v}}

\begin{document}
\hfill KA-TP-12-2001
\begin{center}{{\huge Gauge invariant decomposition of 1-loop multiparticle
scattering amplitudes\\}
\vskip0.2truecm
 {\large Alessandro Vicini\footnote{vicini@particle.uni-karlsruhe.de}\\
Institut f\"ur Theoretische Physik, Universit\"at Karlsruhe, D-76128
Karlsruhe (Germany)\\}}
\end{center}
Abstract: A simple algorithm is presented to decompose any 1-loop amplitude for
scattering processes of the class 2~fermions$\to$ 4~fermions into a
fixed number of gauge-invariant form factors.
The structure of the amplitude is simpler than in the
conventional approaches and its numerical evaluation is made faster.
The algorithm can be efficiently applied also to amplitudes with several 
thousands of Feynman diagrams.
\vskip0.5cm

The complete calculation of the electroweak radiative corrections to the 
class of processes $e^+e^-\to 4 f$ is still missing, for several
reasons of theoretical but also especially technical origin.
At 1-loop the probability amplitude is given by the sum of several
thousands of Feynman diagrams, that we try to organize in a sensible way.
The algebraic programs which should perform the simplifications find severe
obstacles, because they have to deal with huge expressions:
it is indeed very difficult to look for simplification patterns, taking the
amplitude as a whole. We propose the opposite approach:
having a physically motivated structure in mind, we can apply
it systematically to every single Feynman diagram. The simplification of 
a small expression is very efficient and the bookkeeping of the various
contributions follows from the beginning a precise pattern.
The physical picture we are thinking of is the following:
the interaction of elementary fermionic neutral and charged currents,
which are factorized in the amplitude, is described by a rank-3 Lorentz
tensor, which can be evaluated either at tree- or at 1-loop level and
can be decomposed in a gauge-invariant way.

The paper is organized in the following way.
In section 1 the present approaches and their
efficiency in dealing with the scattering amplitudes are briefly described.
In section 2 the decomposition proposed in this paper is 
formulated, proving the gauge invariance of the coefficients.
In section 3 the algorithm to reduce any 1-loop Feynman diagram into the 
proposed form is described, and in section 4 we make some final remarks.

\section{Present approaches}
\label{sec1}
The number of Feynman diagrams which contribute to the probability
amplitude of a process of the class $e^+e^-\to 4f$ is very large.
In the following we consider only processes
with massless external fermions; in table \ref{numberofdiagrams}
we list the number of diagrams which
are part of the virtual corrections to some representative processes,
omitting the tadpoles contributions.
One first comment is that it does not make sense to consider,
pictorially, Feynman diagrams as building blocks of the calculations.
As the well known example of on-shell $W$-pair production shows,
individual Feynman diagrams contain unitarity-violating terms,
which cancel in the sum  at the level of the amplitude;
the interference between different diagrams yields most of the 
physical contributions.

\begin{table}[h]
\begin{tabular}{|l|r|r|r|r|}
\hline
process & diagrams & spinor products & form factors & form factor +
eq. of motion\\
\hline
$e^+e^-\to\mu^-{\overline\nu}_{\mu} u {\overline d}$ & 1907 & 2069 & 64 & 16\\
\hline
$e^+e^-\to e^+e^- d {\overline d}$ & 8522 & 4056 & 128 & 34\\
\hline
$e^+e^-\to e^+ e^- e^+ e^-$ & 21444 & 9157 & 384 & 104\\
\hline
\end{tabular}
\caption{Some representative processes of the class $e^+e^-\to
4f$ and the number of Feynman diagrams due to their 1-loop virtual
corrections, excluding tadpoles contributions, in the limit of massless
external particles.  }
\label{numberofdiagrams}
\end{table}

Since most of the processes of the class $e^+e^-\to 4 f$ are mediated by 
the exchange of a pair of massive vector bosons, a convenient approach
\cite{bbc} is to expand the amplitude about the poles of resonant vector bosons
propagators: one obtains double-resonant, single-resonant and 
non-resonant terms, whose coefficients are gauge invariant, on mathematical
grounds.
This expansion describes the amplitude, considering the 
structure of the resonances of the intermediate states.
On the other hand, a lot of information about the kinematics of the
process is contained in the external fermion lines.
The spinor products which appear in a complete 1-loop calculation are 
very much involved and can not be easily simplified.
We would like to concentrate our attention on this point.

In processes of the class $2f\to 2f$ it is convenient, from the
computational point of view, to square the amplitude and to evaluate the 
traces of the spinor products, to sum over the spin polarizations.
Due to the very large number of diagrams and to the length of the
resulting traces, the same strategy is not so
efficient when we consider processes of the class $2f\to 4f$. In this
case it is common practice to adopt the helicity-amplitude technique,
which consists in the projection of the amplitude on the helicity states, and
then in its numerical evaluation.
To square the amplitude is then just a trivial product of complex
numbers and the unpolarized cross-section is obtained summing
incoherently over all helicity states.
Despite the last sum, the number of operations in the latter approach 
is considerably smaller than the one with the conventional trace technique.

Still, the helicity-state approach is not satisfactory as soon as we
consider  a full 1-loop amplitude.
Each diagram contains the spinors and the fermion lines of the initial 
and final-state fermions, which interact by the exchange of gauge bosons.
To simplify the amplitude, all the internal Lorentz indices are
contracted (e.g. \cite{pittau}).
As a result, each Feynman diagram has a factor proportional to some
spinor products,
of the form
$ S = [\vbar \Gamma_1 u]~~[\ubar \Gamma_2 v]~~[\ubar \Gamma_3 v]$.
The $\Gamma_i$ are products of up to 3 Dirac matrices, contracted with
the momenta flowing in the internal fermion propagators or with those
present in some interaction vertices;
$u$ and $v$ denote the particle and anti-particle spinors.
The large number of Feynman diagrams yields a correspondingly large
number of distinct factors $S$, each of them describing the specific
kinematics of its parent diagram.
Even after a heavy use of the Dirac algebra and of the Fierz identities, 
it is difficult to reduce the number of distinct factors $S$ much below
the number of Feynman diagrams of the process.
The Fierz identities express a redundancy typical of these massless
spinor products. Our aim is to cast all expressions into a form which
is unique with respect to a given basis and which is suitable for a simple
semi-classical interpretation.

In tree-level calculations the problem of the large number of spinor
products has been circumvented by the
observation that there are basic functions \cite{gk},
which describe the spinorial 
part of diagrams with  0 or 1 trilinear gauge boson interaction vertices.
In this way only two basic functions have to be used, where different
diagrams are expressed by appropriate permutations of the arguments of these 
functions.
The same approach could be followed at 1-loop as well, but presents some 
disadvantages. In fact the number of basic spinorial 
functions is larger,
depending on the number of internal propagators in the loop: there are 4 
basic functions with box integrals, 2 with pentagon integrals and 1 with
hexagon integrals; self-energy and vertex corrections respect the
structure of the Born diagrams.
Unfortunately the use of these basic functions does not 
help at all in solving the questions of gauge and unitarity
cancellations:
only some combinations of diagrams are well
behaved in the high energy limit, but the use of basic functions does not
provide an algorithm to find the proper combination of diagrams.

\section{Gauge invariant decomposition of the amplitude}
\label{decomp}
The decomposition proposed in this paper is based on the observation
that each external fermion line should be decomposed in the basis of
the Dirac $\gamma$ matrices. 
Since we consider external massless fermions, only the elements
$\gamma^{\mu}$ and $\gamma^{\mu}\gamma_5$ are relevant.
Any product of 3, 5, 7, ... $\gamma$ matrices can be reduced to a
combination of those two elements of the basis of the algebra.
For convenience we trade these two elements with the right- and
left-handed combinations $\gamma^{\mu}\omega_{\pm}$, where
$\omega_{\pm}=(1\pm\gamma_5)/2$.
For example, we can write
$\ubar ~\aslash \bslash \gamma^{\mu} \cslash \dslash ~v = 
c_1(a,b,c,d) ~[\ubar ~\gamma^{\mu} \omega_+ ~v]~+~
c_2(a,b,c,d) ~[\ubar ~\gamma^{\mu} \omega_- ~v]$
where $c_1$ and $c_2$ are two scalar expressions.
Since we consider massless external fermions, the chiral projections 
$\omega_{\pm} ~v$ coincide with the helicity projections of the spinors.

In this section we present the general structure of the 1-loop
amplitude of processes of the class $e^+e^-\to 4f$.
We present first some specific example, discuss the gauge invariance of the 
coefficients of the decomposition, and then formulate a general rule.\\
{\bf Example 1: $e^+(p_1)~e^-(p_2)\to \mu^-(p_3)~{\overline\nu}_{\mu}(p_4)
~u(p_5) ~{\overline d}(p_6)$}\\
We claim that each helicity amplitude ${\cal M}^{(\lambda)}$,
parametrized by the index $\lambda$ can be written as:
\bea
&&{\cal M}^{(\lambda)}~=~
J_{\alpha}^{12(\lambda)}~J_{\beta}^{34(\lambda)}~
J_{\gamma}^{56(\lambda)}~
\sum_{i=1}^{64} s_i ~w_i^{\alpha\beta\gamma}\\
&&J_{\alpha}^{12(\lambda)}=~\vbar_{\lambda}(p_1)\gamma_{\alpha} u_{\lambda}(p_2),~~~
J_{\beta}^{34(\lambda)}=~
\ubar_{\lambda}(p_3)\gamma_{\beta} v_{\lambda}(p_4),~~~
J_{\gamma}^{56(\lambda)}=~\ubar_{\lambda}(p_5)\gamma_{\gamma} v_{\lambda}(p_6) \nonumber
\eea
The 64 tensors $w^{\alpha\beta\gamma}_i$ are given by all
combinations of 4 independent vectors $q_j^{\mu}$, namely
$q_l^{\alpha}~q_m^{\beta}~q_n^{\gamma}~~(l,m,n=1,\dots,4)$. The
coefficients $s_i$ are scalar coefficients, functions of the external
momenta and of the masses of the internal particles.
Since we assume the vectors $q_j^{\mu}$ to be independent of each other, 
they form a complete basis of tensors of rank-3. We will discuss in the
next section how to choose the momenta $q_j^{\mu}$.
The helicity amplitude $ {\cal M}^{(\lambda)} $ is a physical quantity
and is therefore gauge invariant. 
Neither the external currents $J$ nor the tensors $w_i$ 
depend on the choice of the gauge parameter.
Since we decompose a gauge-invariant quantity in a basis of linearly
independent tensors, each coefficient of this decomposition has to be
separately gauge invariant.\\
{\bf Example 2: $e^+(p_1)~e^-(p_2)\to e^+(p_3)~e^-(p_4)
~d(p_5) ~{\overline d}(p_6)$}\\
In a way analogous to the previous example, we write:
\bea
&&{\cal M}^{(\lambda)}~=~
J_{\alpha}^{12(\lambda)}~J_{\beta}^{34(\lambda)}~
J_{\gamma}^{56(\lambda)}~
\sum_{i=1}^{64} s_i w_i^{\alpha\beta\gamma}~+~
J_{\alpha}^{13(\lambda)}~J_{\beta}^{24(\lambda)}~
J_{\gamma}^{56(\lambda)}~
\sum_{i=1}^{64} t_i w_i^{\alpha\beta\gamma}~~~~\nonumber\\
&&~~~~~~~~~=~
J_{\gamma}^{56(\lambda)}
\sum_{i=1}^{64} w_i^{\alpha\beta\gamma}
\left(
s_i J_{\alpha}^{12(\lambda)}~J_{\beta}^{34(\lambda)}~+~
t_i J_{\alpha}^{13(\lambda)}~J_{\beta}^{24(\lambda)}
\right)\\
&&J_{\alpha}^{12(\lambda)}~=~\vbar_{\lambda}(p_1)\gamma_{\alpha} u_{\lambda}(p_2),~~
J_{\beta}^{34(\lambda)}~=~
\ubar_{\lambda}(p_3)\gamma_{\beta} v_{\lambda}(p_4),~~
J_{\gamma}^{56(\lambda)}~=~\ubar_{\lambda}(p_5)\gamma_{\gamma} v_{\lambda}(p_6) 
\nonumber\\
&&J_{\alpha}^{13(\lambda)}~=~\vbar_{\lambda}(p_1)\gamma_{\alpha} v_{\lambda}(p_3),~~
J_{\beta}^{24(\lambda)}~=~
\ubar_{\lambda}(p_4)\gamma_{\beta} u_{\lambda}(p_2) \nonumber
\eea
As in the previous example, the coefficient of each tensor $w_i$ has to 
be separately gauge invariant.
With only few exceptions in the phase space, the tensors
$J_{\alpha}^{12(\lambda)}~J_{\beta}^{34(\lambda)}$ and
$J_{\alpha}^{13(\lambda)}~J_{\beta}^{24(\lambda)}$ are independent of
each other.
Therefore the two coefficients $s_i$ and $t_i$ have to be separately
gauge-invariant, in order to make a combination gauge-invariant as well.
The coefficients $s_i$ and $t_i$
are separately gauge-invariant, even when the two tensors
$J_{\alpha}^{12(\lambda)}~J_{\beta}^{34(\lambda)}$ and
$J_{\alpha}^{13(\lambda)}~J_{\beta}^{24(\lambda)}$ are linearly
dependent, because the functional dependence which guarantees the gauge
cancellation remains the same throughout the whole phase-space.\\
{\bf Example 3: $e^+(p_1)~e^-(p_2)\to e^+(p_3)~e^-(p_4)
~e^+(p_5) ~e^-(p_6)$}
\bea
&&{\cal M}^{(\lambda)}~=~
\sum_{i=1}^{64} w_i^{\alpha\beta\gamma}
\Big(
s^{(1)}_i J_{\alpha}^{12(\lambda)}~J_{\beta}^{34(\lambda)}~J_{\gamma}^{56(\lambda)}~+~
s^{(2)}_i J_{\alpha}^{12(\lambda)}~J_{\beta}^{36(\lambda)}~J_{\gamma}^{45(\lambda)}~+~
t^{(1)}_i
J_{\alpha}^{13(\lambda)}~J_{\beta}^{24(\lambda)}~J_{\gamma}^{56(\lambda)}
~+~\nonumber \\
&&~~~~~~~~~~~~~~~~~~~~~~~~~~
t^{(2)}_i J_{\alpha}^{13(\lambda)}~J_{\beta}^{26(\lambda)}~J_{\gamma}^{45(\lambda)} ~+~
u^{(1)}_i J_{\alpha}^{15(\lambda)}~J_{\beta}^{24(\lambda)}~J_{\gamma}^{36(\lambda)}~+~
u^{(2)}_i J_{\alpha}^{15(\lambda)}~J_{\beta}^{26(\lambda)}~J_{\gamma}^{34(\lambda)}
\Big)
\eea
The structure of this formula derives from the presence of identical
particles in the initial and in the final state.
Each expression in round brackets has to be separately gauge-invariant.
As in the previous example, we observe that for generic points in the
phase space the 6 different tensors $J_{\alpha} J_{\beta} J_{\gamma}$ 
are independent of each other; therefore the dependence on the gauge
parameter has to cancel separately in each coefficient $s,~t,~u$.\\
{\bf General rule}\\
Each Feynman diagram can be reduced, with the algorithm proposed in the next
section, to one current product,
which has initial and final state fermions connected in the same way as
the external fermion lines of the diagram.
Each diagram contributes to the 64 coefficients of the tensors
$w_i^{\alpha\beta\gamma}$ of its specific current product.
Since the current products are in general independent of each other, the 
proof of the gauge-invariance of the scalar coefficients holds for each
of them separately.

\section{Reduction algorithm}
\label{alg}

The algorithm to reduce any external fermion line to a current $J_{\alpha}$ 
works in 4 dimensions and might have some troubles in dimensional
regularization, because of the $\gamma_5$ problem.
We regularize soft-infrared and collinear divergences by means of photon
and fermion masses. Tree level, virtual 1-loop diagrams with box,
pentagon and hexagon loop integrals are ultraviolet finite and can be
manipulated in 4 dimensions. Self-energy and vertex corrections can be
considered after renormalization: namely our approach can be applied
once the limit $n\to 4$ has been taken, $n$ being the number of
space-time dimensions.

{\bf 1.} Since we consider massless external fermions, each external fermion
line contains the product of an odd number of $\gamma$ matrices. We
reduce these products in terms of one single $\gamma$ matrix, by
repeated use of the Chrisolm identity:
\be
\label{Chrisolm}
\gamma^{\alpha}\gamma^{\beta}\gamma^{\gamma}~=~
\left(
g^{\alpha\beta} g^{\gamma\delta}-
g^{\alpha\gamma} g^{\beta\delta}+
g^{\alpha\delta} g^{\beta\gamma}+
i \gamma_5 \varepsilon^{\alpha \beta\gamma\delta}
\right)\gamma_{\delta}~\equiv~
\left( S^{\alpha\beta\gamma\delta} +
i \gamma_5 \varepsilon^{\alpha \beta\gamma\delta}\right)\gamma_{\delta}
~\equiv~X^{\alpha\beta\gamma\delta}\gamma_{\delta}
\label{chri3}
\ee
The case with 5 matrices is obtained in 2 steps and gives:
\bea
&&\gamma^{\alpha}~\gamma^{\beta}~\gamma^{\gamma}~\gamma^{\delta}~\gamma^{\zeta}
~=\label{chri5}\\
&&\Big[ \left(
     \varepsilon^{\gamma \delta \zeta \mu} g^{\alpha \beta} - 
     \varepsilon^{ \beta \delta \zeta \mu} g^{\alpha \gamma} + 
     \varepsilon^{ \alpha \delta \zeta \mu} g^{\beta \gamma} + 
     \varepsilon^{ \alpha \beta \gamma \mu} g^{\delta \zeta} - 
     \varepsilon^{ \alpha \beta \gamma \zeta} g^{\delta \mu} + 
     \varepsilon^{ \alpha \beta \gamma \delta} g^{\zeta \mu}
     \right) i~\gamma_5 + \nonumber \\
&&~~g^{\alpha \mu}\,g^{\beta \zeta}\,g^{\gamma \delta} + 
  g^{\alpha \zeta}\,g^{\beta \mu}\,g^{\gamma \delta} + 
  g^{\alpha \mu}\,g^{\beta \delta}\,g^{\gamma \zeta} - 
  g^{\alpha \delta}\,g^{\beta \mu}\,g^{\gamma \zeta} - 
  g^{\alpha \zeta}\,g^{\beta \delta}\,g^{\gamma \mu} + \nonumber \\
&&~~g^{\alpha \delta}\,g^{\beta \zeta}\,g^{\gamma \mu} + 
  g^{\alpha \mu}\,g^{\beta \gamma}\,g^{\delta \zeta} - 
  g^{\alpha \gamma}\,g^{\beta \mu}\,g^{\delta \zeta} + 
  g^{\alpha \beta}\,g^{\gamma \mu}\,g^{\delta \zeta} - 
  g^{\alpha \zeta}\,g^{\beta \gamma}\,g^{\delta \mu} + \nonumber \\
&&~~g^{\alpha \gamma}\,g^{\beta \zeta}\,g^{\delta \mu} - 
  g^{\alpha \beta}\,g^{\gamma \zeta}\,g^{\delta \mu} + 
  g^{\alpha \delta}\,g^{\beta \gamma}\,g^{\zeta \mu} - 
  g^{\alpha \gamma}\,g^{\beta \delta}\,g^{\zeta \mu} + 
  g^{\alpha \beta}\,g^{\gamma \delta}\,g^{\zeta \mu}  \Big]
   \gamma_{\mu} \nonumber
\eea
We observe that for processes of the class $e^+e^-\to 4f$, lines with a 
product of 7 $\gamma$ matrices appear only in the form
$\gamma_{\mu}
\gamma^{\alpha}~\gamma^{\beta}~\gamma^{\gamma}~\gamma^{\delta}~\gamma^{\zeta}
\gamma^{\mu}~=~-2~
\gamma^{\zeta}~\gamma^{\delta}~\gamma^{\gamma}~\gamma^{\beta}~\gamma^{\alpha}
$.
We apply substitutions \equ{chri3} and \equ{chri5} to the 3 fermion
lines of a Feynman diagram belonging to our class of processes
and observe that after the contraction of all Lorentz indices
the result has the structure
$J_{\alpha} J_{\beta} J_{\gamma} R^{\alpha\beta\gamma}$;
the tensor $R^{\alpha\beta\gamma}$ still contains metric and
Levi-Civita tensors and the external momenta.

{\bf 2.}
We choose 4 vectors $q_j^{\mu}$ independent of each 
other and write
\bea
&&g^{\mu\nu}=\frac{1}{a {\overline a}}
\sum_{i,j} \left( a_i\cdot a_j\right) q_i^{\mu} q_j^{\nu}\label{met}\\
&&\varepsilon^{\alpha r s t}=
\frac{1}{a}
\left\{
\left(a_{1\lambda}\cdot \varepsilon^{\lambda r s t}\right)q_1^{\alpha}~+~
\left(a_{2\lambda}\cdot \varepsilon^{\lambda r s t}\right)q_2^{\alpha}~+~
\left(a_{3\lambda}\cdot \varepsilon^{\lambda r s t}\right)q_3^{\alpha}~+~
\left(a_{4\lambda}\cdot \varepsilon^{\lambda r s t}\right)q_4^{\alpha}
\right\}\label{epsq}\\
&&a=q_{1\alpha} q_{2\beta} q_{3\gamma} q_{4\delta}~
\varepsilon^{\alpha\beta\gamma\delta} ~\equiv~\varepsilon^{q_1 q_2 q_3 q_4}, 
~~{\overline a}=\varepsilon_{q_1 q_2 q_3 q_4},\nonumber\\
&&a_1^{\mu}=\varepsilon^{\mu q_2 q_3 q_4},~~
a_2^{\mu}=\varepsilon^{q_1 \mu q_3 q_4},~~
a_3^{\mu}=\varepsilon^{q_1 q_2 \mu q_4},~~
a_4^{\mu}=\varepsilon^{q_1 q_2 q_3 \mu}\nonumber
\eea
The product $a {\overline a}$ is the (Gram-) determinant of the matrix
$q_i \cdot q_j$, while the products $a_i\cdot a_j$ yield the minors of
the latter.
We notice the the contraction of two Levi-Civita tensors yields a
combination of scalar products.
Equation (\ref{epsq}) is the dual of the Schouten identity and expresses an
axial vector in terms of polar vectors. The quantity $a$ is
pseudo-scalar with respect to parity transformations.
Using the substitution rules \equ{met} and \equ{epsq}
we are able to express all
the metric and Levi-Civita tensors which appear in
$R^{\alpha\beta\gamma}$ in terms of the vectors 
$q_j^{\alpha}$. The tensor $R^{\alpha\beta\gamma}$ contains now only
combinations of the vectors $q_j$ and of the external momenta.

{\bf 3.}
In order the decompositions \equ{met} and \equ{epsq} to hold, we need 4
independent Lorentz-vectors, so that the pseudo-scalars $a, {\overline
a}$ do not vanish. It is convenient to identify the $q_j^{\mu}$ with
some of the external momenta, to avoid cumbersome algebraic expressions
and to exploit the equation of motion of the external spinors.
It is important to check that the chosen momenta remain independent in
the whole phase-space, to avoid artificial numerical instabilities.
In Example 1 the choice $q_1 = p_3,~~q_2 = p_4,~~q_3 = p_5,~~q_4 = p_6$
is valid everywhere except when both $W's$ are at rest.
In the latter case it is possible to have configurations in which 
the decay fermions of one $W$ are parallel to those of the second $W$.
On the other hand it is important to observe that such configurations
form a negligible fraction of the phase-space and 
that the amplitude has a much simpler expression, thanks to the reduced
number of final state momenta. Also these configurations can therefore
be separately, easily, evaluated.

In the problem we have 6 external momenta $p_i$. By means of
energy-momentum conservation we are left with 5 momenta, for instance
we write $p_1 = -p_2+p_3+p_4+p_5+p_6$.
Observing that
a 4-dimensional space-time is completely spanned by 4 independent
momenta, we decompose one momentum in terms of the other four, in our
example $p_2 = c_{23} p_3 +c_{24} p_4 +c_{25} p_5 +c_{26} p_6$.
With this choice for the vectors $q$, the tensor $R^{\alpha\beta\gamma}$ 
contains only factors $p_l^{\alpha} p_m^{\beta}
p_n^{\gamma},~~(l,m,n=3,\dots,6)$, which form (with the {\it caveat}
mentioned above) a complete basis of rank-3 Lorentz tensors, but not
metric or Levi-Civita tensors.
With this choice for the $q$'s, the number of terms of the decomposition is
reduced to 16 in example 1, to 34 in example 2 and to
104 in example 3.\\
A more solid, but less practical, solution consists in choosing
$q_1=(1,0,0,0),~~q_2=(0,1,0,0),~~q_3=(0,0,1,0),~~q_4=(0,0,0,1)$.
In this way, equations \equ{met} and \equ{epsq} always hold, but we are
forced to decompose all the external momenta along this special basis.

\section{Discussion}
1) The proposed decomposition organizes the probability amplitude in a
fixed number of terms: depending on the final state this number ranges
from 64 to 384 and can be further reduced by means of the equations of
motion.
This number has to be compared with the the one of
Feynman diagrams of the corresponding reactions, or with the number of
spinor products which result, after simplification, in the conventional
helicity amplitude approach; they are both more than one order of
magnitude larger (cf. table \ref{numberofdiagrams}).
This number is fixed {\it a priori} and depends only on the number of
identical particles in initial and final states.
In particular, it is not related at all with the number of Feynman
diagrams of the reaction.

2) Each coefficient in the decomposition can be evaluated at the lowest
order, or including the 1-loop corrections. The latter can be expanded,
if wished, according to the pole expansion.

3) The coefficients of the decomposition are by construction gauge
invariant. This property enforces naturally, in the symbolic manipulation, 
some of the expected gauge cancellations, because the terms which have
to cancel are collected together by the decomposition algorithm.
As a consequence, the numerical evaluation of the coefficients is more
stable and accurate.

4) The size of the amplitude remains big, as compared with the usual
approach, but its structure is much simpler. 
The factorization of the currents reflects a semi-classical behaviour of 
the amplitude, on top of which the radiative corrections act.
This factorization is not trivial, because of the presence of fermion
lines with up to 7 $\gamma$ matrices.
By semi-classical behaviour we mean that the current products
give a simple and intuitive pictorial description of the
fermion and flavour flows.
Even if it is possible to reduce further the products
$J_{\alpha} J_{\beta} J_{\gamma} w^{\alpha\beta\gamma}$ in terms of
simpler spinor products of the form ${\overline u}(p_i) v(p_j)$, the
proposed decomposition seems preferable, precisely because it does not
mix particles with different flavour. In our decomposition 
the scalar coefficients express the weight of a given configuration
of physical currents.

5) The algorithm is very efficient, because it works on one
Feynman diagram at a time, with a simple set of substitution rules.
Collecting together the contributions to a given scalar coefficient
does not create problems from the computational point of view,
even in the case of several thousands of diagrams.

6) The amount of CPU-time needed to evaluate the amplitude is reduced
with respect to the normal helicity amplitude approach.
In fact the number of spinor
products which appear in the final expression is much smaller than in
the previous case;
the algebraic complexity induced by the rules
\equ{chri3},~\equ{chri5},~\equ{met} and \equ{epsq} does not cost
a relevant increase of the CPU-time, because the number of scalar
products between the external momenta remains exactly the same.
In the case of $e^+e^-\to \mu^- {\overline\nu}_{\mu} u {\overline d}$
this saving can be quantified to be approximately 15\%.
This number is not negligible, if one considers that the evaluation of
one point of total cross-section requires about 1 week, on a 600 MHz
Digital Sigma Station. 

7) The introduction of the full set of electroweak 1-loop corrections
into event generators seems to be not viable, as long as the
CPU-time needed to evaluate the squared matrix elements in 1 point of
the phase-space is of the order of 1 second. 
The only realistic alternative is to perform a careful evaluation of the 
differential cross-section and then to fit the latter using some
appropriate function.
Although the scalar coefficients $s,~t,~u$ are not observable quantities, 
they give a weight to the different kinematical configurations of the external
currents. 
This relation to physical quantities, supported by their
gauge-invariance, indicates that they are good
candidates to become the building blocks of the calculation,
instead of the Feynman diagrams.
Using, for instance, a function organized as in the pole expansion,
it should be possible to fit very accurately each scalar coefficient.


\section{Conclusions}
In this paper we have presented a simple algorithm to organize any
1-loop scattering amplitude of processes of the class $e^+e^-\to 4f$
in a physically motivated way: we decompose it in gauge invariant
form factors which are suitable for analytical checks or for numerical
evaluation and subsequent fits.
A package implementing this algorithm has been written and tested \cite{fc}.

As a first application the 1-loop amplitude for the
process $e^+e^-\to \mu^-{\overline\nu}_{\mu} u {\overline d}$ will be
presented in a 
forthcoming publication, its detailed description being well beyond the
scope of this paper.\\

\section{Acknowledgments}
We thank O. Brein, T. Hahn, W. Hollik and D. St\"ockinger 
for carefully reading the manuscript and useful discussions.
This work was supported by the Deutsche Forschungsgemeinschaft
(Forschergruppe ``Quantenfeldtheorie, Computeralgebra und Monte Carlo
Simulation'').


\end{document}